\documentstyle[12pt]{laa}
\def\ltsima{$\; \buildrel < \over \sim \;$}
\def\lsim{\lower.5ex\hbox{\ltsima}}
\def\gtsima{$\; \buildrel > \over \sim \;$}
\def\gsim{\lower.5ex\hbox{\gtsima}}

\def\aa #1 #2 {A\&A, {#1}, #2}
\def\aas #1 #2 {A\&AS, {#1}, #2}
\def\araa #1 #2 {ARA\&A, {#1}, #2}
\def\mon #1 #2 {MNRAS, {#1}, #2}
\def\apj #1 #2 {ApJ, {#1}, #2}
\def\apjs #1 #2 {ApJS, {#1}, #2}
\def\apjl #1 #2 {ApJ, {#1}, #2}
\def\aj #1 #2 {AJ, {#1}, #2}
\def\nat #1 #2 {Nature, {#1}, #2}
\def\pasj #1 #2 {PASJ, {#1}, #2}
\def\pasp #1 #2 {PASP, {#1}, #2}

\input psfig.tex
                                                                              
\begin{document}
 \thesaurus{
             (08.14.1:  
              08.16.6)} 

\title{Isolated neutron stars, their $\gamma$-ray efficiencies and EGRET observations}
\author{Paolo Goldoni\inst{1,2} and Carlo Musso\inst{1}}
\institute{$^1$ Istituto di Fisica Cosmica, CNR, v. Bassini 15, I-20133, Milano,
Italy \\ $^2$ Dipartimento di Fisica, Universit\`a di Milano, v. Celoria 16, I-20133, Milano, Italy}
\date{\today}
\maketitle

\begin{abstract} 

We examine a sample of detected and undetected Isolated Neutron Stars
(INSs), selected on the basis of their energy loss and distance, in order to
maximize their detection probability. Since the sample we consider
encompasses more upper limits than detections, we make use, for the first
time in this field of astronomy, of survival analysis procedures through
the ASURV software package (Feigelson \& Nelson 1985, Isobe et al. 1986).
We show that these techniques lead to a better understanding of the physical
processes at work in high-energy emission from INSs. In particular, the
recent detection of PSR 1951+32 and upper limits from pulsars pointed but
not detected by EGRET show that the $\gamma $-ray efficiency of ISN is not
correlated to any simple pulsar parameter.

\end{abstract}

\section {Introduction}

After the recent PSR 1951+32 detection (Ramanamurthy et al. 1995a), six INSs
have been observed as high-energy $\gamma $-ray sources by EGRET, at 
energies above hundreds of $MeV$. One should also consider a preliminary
report (Ramanamurthy et al. 1995b, these proceedings) of an EGRET detection
of PSR 0656+14. As discussed in a recent paper (Goldoni et al. 1995), before
these last results, the efficiency $\eta_{\gamma}$ of transforming rotational
energy loss into $\gamma $-ray radiation of the first 5 high-energy INSs
seem to grow with the decrease of the surface magnetic field $B$ and of the
period derivative $\dot P$, and with the increase of the age $\tau $
(Harding 1981;Thompson et al. 1994) and of the period $P$.

As we will show, PSR 1951+32 detection rules out linear correlation between
$\eta_{\gamma}$ and some pulsar parameter. However, it is worth noting that
the same conclusion could be achieved looking at EGRET upper limits
(Ulmer \& Schroeder 1994), and even at COS-B data (Buccheri et al. 1978).

The complete coverage of the sky by EGRET produced a lot of $\gamma $-ray
upper limits (Thompson et al. 1994) for known radio pulsars in the
Princeton catalog (Taylor et al. 1993). In order to correctly evaluate the
observational data, it is necessary to take into account detections and
non-detections in a self-consistent way (Feigelson et al. 1986): in fact
the information involved in a {\sl non-detected} source is different from
the information coming from a {\sl non-pointed} one.

We discuss a sample of possibly detectable INSs on the basis of their
expected energy flux at Earth $\dot E /d^2$ (Mereghetti et al. 1994,
Thompson et al. 1994). This sample includes in the first place all
EGRET pulsars with the exception of PSR 1055-52, and all ROSAT pulsars.

\section {Selection of the sample}

A key problem is to find a statistically unbiased sample of INSs. This is not
straightforward due to the low number of present day high-energy detections
and to the difficulties of a complete radio pulsar survey (see e.g. Curtis
Michel 1991). We excluded millisecond pulsars as no one of them was detected
as high-energy radiation emitter (Fierro et al. 1995), neither they were detected
as a whole in clusters (Barret et al. 1993; Michelson et al. 1994).

We selected the first 40 INSs by their $\dot E /d^2$ ranking. This criterion
selects young nearby objects. As it is shown in Table 1, it works well for the
first four EGRET detections and even for PSR 1951+32, but it fails to predict
the detection of PSR 1055-52. So one could expect not only high-rank INSs to
be detectable, but even objects in a lower position (Hermsen et al. 1993).
We took the beaming factor constant (1/4$\pi$), considering its variation as
a second order effect (Helfand 1994). We excluded farther objects by simple
energetic arguments, e.g. their $\gamma $-ray luminosity would be greater
than their $\dot E$ with any reasonable beaming choice. The properties of
the sources are listed in Table 1.


\begin{table*}
\[
\begin{array}{cr|c|c|c|c|c|c|c|c|c|c}
\hline
\noalign{\smallskip}
& {\rm Source} & {\rm Period} & {\rm ~Log~\dot P~} & {\rm Log~\tau} & {\rm Log~B} & {\rm Log~\dot E} & {\rm Distance} & {\rm Log~\dot E/ d^2} & {\rm Photon~flux} & {\rm Energy~flux} & {\rm Log~\eta_{\gamma}} \\ & & s & s/s & ys & gauss & erg/s & kpc & erg

/cm^2/s & 10^{-7}~ph/cm^2/s & 10^{-10}~erg/cm^2/s & \\
\noalign{\smallskip}
\hline
\noalign{\smallskip}
{\rm X} & {\bf CRAB}        & 0.033 & -12.38 & 3.1 & 12.57 & 38.65 &    2 &  -4.91 &   18 &   11 &  -4.05 \\
{\rm X} & {\bf VELA}        & 0.089 & -12.95 & 4.1 & 12.50 & 36.84 &  0.5 &  -5.51 &   78 &   74 &  -2.62 \\
{\rm X} & {\bf GEMINGA}     & 0.237 & -13.93 & 5.5 & 12.23 & 34.51 &  0.1 &  -6.79 &   29 &   37 &  -0.82 \\
{\rm X} & {\bf PSR 1706-44} & 0.102 & -12.99 & 4.2 & 12.51 & 36.53 &  1.8 &  -6.94 &   10 &  8.9 &  -2.11 \\
{\rm X} & {\rm PSR}~1509-58 & 0.150 & -11.82 & 3.2 & 13.18 & 37.25 &  4.4 &  -6.99 & <5.5 & <3.6 & <-2.50 \\
{\rm X} & {\bf PSR 1951+32} & 0.039 & -14.23 & 5.0 & 11.69 & 36.57 &  2.5 &  -7.18 &  1.6 &  1.7 &  -2.60 \\
        & {\rm PSR}~1046-58 & 0.124 & -13.02 & 4.3 & 12.54 & 36.30 &  3.0 &  -7.60 & <5.8 & <3.1 & <-1.64 \\
{\rm X} & {\rm PSR}~1823-13 & 0.101 & -13.13 & 4.3 & 12.45 & 36.45 &  4.1 &  -7.73 & <4.1 & <4.1 & <-1.86 \\
{\rm X} & {\rm PSR}~1800-21 & 0.134 & -12.87 & 4.2 & 12.63 & 36.35 &  3.9 &  -7.80 & <4.7 & <4.7 & <-1.75 \\
{\rm X} & {\rm PSR}~1929+10 & 0.227 & -14.94 & 6.5 & 11.72 & 33.59 &  0.2 &  -7.82 & <1.6 & <2.6 & <-1.81 \\
        & {\rm PSR}~1757-24 & 0.125 & -12.89 & 4.2 & 12.61 & 36.41 &  4.6 &  -7.87 & <2.8 & <2.7 & <-1.71 \\
        & {\rm PSR}~1727-33 & 0.139 & -13.07 & 4.4 & 12.54 & 36.09 &  4.2 &  -8.12 & <3.0 & <3.2 & <-1.39 \\
{\rm X} & {\rm PSR}~0656+14 & 0.385 & -13.26 & 5.0 & 12.67 & 34.58 &  0.8 &  -8.13 & <1.2 & <1.4 & <-1.88 \\
{\rm X} & {\rm PSR}~0540-69 & 0.050 & -12.32 & 3.2 & 12.70 & 38.17 & 49.4 &  -8.17 & <1.3 & <0.8 & <-2.10 \\
{\rm X} & {\rm PSR}~0114+58 & 0.101 & -14.23 & 5.4 & 11.89 & 35.35 &  2.1 &  -8.26 & <2.4 & <3.8 & <-1.17 \\
        & {\rm PSR}~0740-28 & 0.167 & -13.77 & 5.2 & 12.23 & 35.16 &  1.9 &  -8.35 & <1.6 & <2.2 & <-1.30 \\
        & {\rm PSR}~1853+01 & 0.267 & -12.68 & 3.3 & 12.88 & 35.63 &  4.3 &  -8.36 & <7.0 & <4.1 & <-0.80 \\
{\rm X} & {\rm PSR}~0950+08 & 0.253 & -15.64 & 7.2 & 11.39 & 32.75 &  0.1 &  -8.44 & <1.5 & <7.2 & <-1.14 \\
        & {\rm PSR}~1610-50 & 0.232 & -14.31 & 3.9 & 13.03 & 36.19 &  7.3 &  -8.48 & <3.8 & <3.1 & <-1.02 \\
        & {\rm PSR}~1338-62 & 0.193 & -12.59 & 4.1 & 12.85 & 36.14 &  8.7 &  -8.69 & <3.4 & <3.1 & <-0.83 \\
        & {\rm PSR}~1830-08 & 0.085 & -14.04 & 5.2 & 11.95 & 35.77 &  5.7 &  -8.70 & <3.5 & <4.8 & <-0.61 \\
{\rm X} & {\bf PSR 1055-52} & 0.197 & -14.21 & 5.7 & 12.04 & 34.48 &  1.5 &  -8.85 &  2.4 &  4.4 &  -0.51 \\
        & {\rm PSR}~0906-49 & 0.107 & -13.82 & 5.0 & 12.11 & 35.69 &  6.6 &  -8.90 & <2.0 & <2.5 & <-0.67 \\
{\rm X} & {\rm PSR}~0355+54 & 0.156 & -14.36 & 5.7 & 11.92 & 34.66 &  2.1 &  -8.93 & <3.0 & <5.3 & <-0.54 \\
{\rm X} & {\rm PSR}~2334+61 & 0.495 & -12.72 & 4.6 & 12.99 & 34.80 &  4.6 &  -8.94 & <4.1 & <3.2 & <-0.60 \\
        & {\rm PSR}~1930+22 & 0.144 & -13.24 & 4.6 & 12.47 & 35.88 &  9.8 &  -9.06 & <3.6 & <4.0 & <-0.34 \\
        & {\rm PSR}~1737-30 & 0.607 & -12.33 & 4.5 & 13.23 & 34.92 &  6.8 &  -9.07 & <3.3 & <0.7 & <-0.42 \\
        & {\rm PSR}~1643-43 & 0.232 & -12.95 & 4.5 & 12.71 & 35.55 &  6.8 &  -9.07 & <4.9 & <5.3 & <-0.21 \\
        & {\rm PSR}~1449-64 & 0.180 & -14.56 & 6.0 & 11.85 & 34.27 &  1.8 &  -9.21 & <2.5 & <5.1 & <-0.11 \\ 
        & {\rm PSR}~1634-45 & 0.119 & -14.50 & 5.8 & 11.79 & 34.88 &  3.8 &  -9.24 & <7.0 &  <12 & <+0.34 \\
        & {\rm PSR}~1719-37 & 0.236 & -13.97 & 5.5 & 12.21 & 34.51 &  2.5 &  -9.25 & <3.0 & <5.0 & <-0.06 \\
        & {\rm PSR}~1702-19 & 0.299 & -14.38 & 6.1 & 12.05 & 33.79 &  1.2 &  -9.32 & <1.1 & <2.2 & <-0.34 \\
        & {\rm PSR}~1822-09 & 0.769 & -13.28 & 5.4 & 12.81 & 33.66 &  1.0 &  -9.32 & <3.8 & <6.1 & <+0.08 \\
        & {\rm PSR}~0450+55 & 0.341 & -14.63 & 6.4 & 11.96 & 33.37 &  0.8 &  -9.37 & <1.5 & <3.1 & <-0.13 \\
        & {\rm PSR}~1221-63 & 0.216 & -14.30 & 5.8 & 12.02 & 34.28 &  2.3 &  -9.39 & <2.4 & <4.2 & <+0.02 \\
        & {\rm PSR}~1356-60 & 0.127 & -14.20 & 5.5 & 11.96 & 35.08 &  5.9 &  -9.42 & <7.4 &  <12 & <+0.49 \\
        & {\rm PSR}~0540+23 & 0.246 & -13.81 & 5.4 & 12.29 & 34.61 &  3.5 &  -9.44 & <1.3 & <2.3 & <+0.07 \\
        & {\rm PSR}~1754-24 & 0.234 & -13.89 & 4.2 & 12.25 & 34.60 &  3.5 &  -9.44 & <3.4 & <5.3 & <+0.17 \\
        & {\rm PSR}~1607-52 & 0.183 & -14.29 & 5.7 & 11.99 & 34.53 &  3.3 &  -9.48 & <3.7 & <6.6 & <+0.29 \\
        & {\rm PSR}~0611+22 & 0.335 & -13.22 & 4.9 & 12.65 & 34.80 & 4.7 &-9.50 & <4.5 & < 5.5 & <+0.26 \\ 
\end{array} \]

\caption{Parameters of the first 40 INSs, ordered by $\dot E/d^2$ ranking 
(Taylor et al. 1993). The "X" labels INSs detected by ROSAT. Waiting for 
a confirmation, for PSR 0656+14 we consider only the upper limit. Fluxes and
efficiencies are derived from $E>100~MeV$ EGRET data (Thompson et al.
1994). The photon flux limits are 99.9 \% confidence limits based on
spatial analysis. The energy flux limits are obtained as in Thompson 1994.} 
\label{Table1} 
\end{table*}

\section{Data Analysis}

An extensive field of statistics, called {\sl survival analysis}, has been
developed to deal with censored data, for which only incomplete information
is present. It has been widely used over some decades in epidemiology and
industrial reliability. In the past years several astronomers (Avni et al.
1980, Feigelson \& Nelson 1985, Schmitt et al. 1985) applied these methods
to samples containing upper limits. In this way it is possible to use each
detection and upper limit in an efficient and well defined manner, minimizing
error sources. In fact, spurious correlations can appear if upper limits are
neglected from the analysis of a sample (Elvis et al. 1981; Feigelson \& Berg
1983). Simulations by Isobe et al. (1986) show that they disappear with a proper
use of existing upper limits. For a complete description of survival analysis,
see Feigelson et al. (1986), where a general bibliography as well as a discussion
of astrophysical problems are presented.

To perform this analysis we used the ASURV Rev 1.2 software package (Isobe \&
Feigelson 1990; La Valley et al. 1992), kindly provided to us by E. B.
Feigelson. The validity of these methods relies on random censoring: the
distribution of upper limits must be independent from the distribution of true
data values; in other words, high and low flux objects must have the same
distribution of observational sensitivity. Censoring becomes more random if
one uses not flux but intrinsic quantities such as luminosity or spectral index,
which are folded with other parameters like distance (Magri et al. 1988).

However, often real observing changes all of this. When an observer does not
detect an object, he usually lowers the threshold by further observations. In
our case this means summing more EGRET observation periods to achieve higher
sensitivities. The example of PSR 1951+32 detection illustrates this process
very well, showing that three years of observations were necessary to perform
this task. We also note that the data reduction method of Ramanamurthy et al.
(1995a) is different from the standard one, employing a new photon selection
method. The same can be said for PSR 0656+14 preliminary detection which was
obtained with a completely different method.

This shows that there is usually no precise flux limit between detections and
non-detections. In the idealized case the sample is formed by detections above a
certain level with non-detections below that level. In our case we will consider
that, thanks to the great EGRET field of view, all the sky received an almost uniform
coverage. We can so define in a satisfactory way an upper limit of detectable photon
flux of $\sim 10^{-7}~ph/cm^2/s$. This limit is more or less coincident with the
$\dot E /d^2$ condition chosen for the INSs' sample. Here we use a few different
tests to establish a correlation between INS's $\gamma $-ray emission and the rotational
parameters. First of all we compare the result of simple correlation tests on the
detected sources; in a second time we include the upper limits in the test procedure.

It should be noted that statistical tests of survival analysis do not consitute a
well defined procedure in astronomical data analysis. This is especially true for
linear fitting techniques which are seldom used in other fields of research. For
this reason we did not follow completely La Valley et al. (1992) advice of using
all the available tests, but we excluded Schmitt's linear regression method
(Schmitt et al. 1985) due to unclear and somewhat arbitrary bin selection. We perform
as a first step the linear correlation tests, and then apply the linear regression
tests. The second step will be mainly useful to show the effects of upper limit
inclusion in our analysis.

\section{Results}

We first made correlation analysis on the detected sources, and then to the whole
sample with the upper limits. We present the results of the same tests even on
ROSAT-detected INSs (Table 2), and we found no difference with the results of the
whole sample. While in the case of detected sources a correlation probability higher
than 95\% is shown by $P$, $\dot P$ and $\tau $ (not by $B$), when we consider the
whole sample (or the X-ray sample), only the age $\tau $ remains above the 95\% threshold.


\begin{table}
\[
\begin{array}{r|c|c|c|c|c}
\hline
\noalign{\smallskip}
\rm Sample & \rm ~Test~ & \rm Period & \rm \dot P & \tau & \rm B \\
\noalign{\smallskip}
\hline
\noalign{\smallskip}
\rm EGRET & (1) & ~~~~0.97~~~~ & ~~~~0.95~~~~ & ~~~~0.99~~~~ & ~~~~0.64~~~~ \\ 
\rm detections & (2) & 0.96 & 0.91 & 0.99 & 0.81 \\
\noalign{\smallskip}
\hline
\noalign{\medskip}
\rm X-ray & (1) & 0.84 & 0.91 & 0.95 & 0.78 \\ 
\rm INSs & (2) & 0.85 & 0.90 & 0.96 & 0.85 \\
\noalign{\smallskip}
\hline
\noalign{\medskip}
\dot E/d^2 & (1) & 0.84 & 0.94 & 0.97 & 0.82 \\
\rm sample & (2) & 0.86 & 0.90 & 0.97 & 0.86 \\
\noalign{\smallskip}
\hline
\end{array}
\]
\caption{Correlation test results for Cox regression (1) and Kendall's tau (2) tests (see Feigelson et al. 1986). The quoted numbers represent the probability that a correlation between the $\gamma $-ray efficiency and the various rotational parameters does exist.}
\label{Table2}
\end{table}

In Fig.1 there are plotted the $\gamma $ efficiencies as a function of $B$ and
$\tau $, for the six sources detected to date and the EGRET upper limits for
other sources in the sample. The linear fits for the first five detected sources
(solid line) and for the whole sample (dashed line) are also drawn. It is apparent
that the detection of PSR 1951+32 and the introduction of the upper limits greatly
affect the $B$ plot and also the $\tau $ plot, reducing the likelihood of the fits.

\begin{figure}

\centerline{\psfig{figure=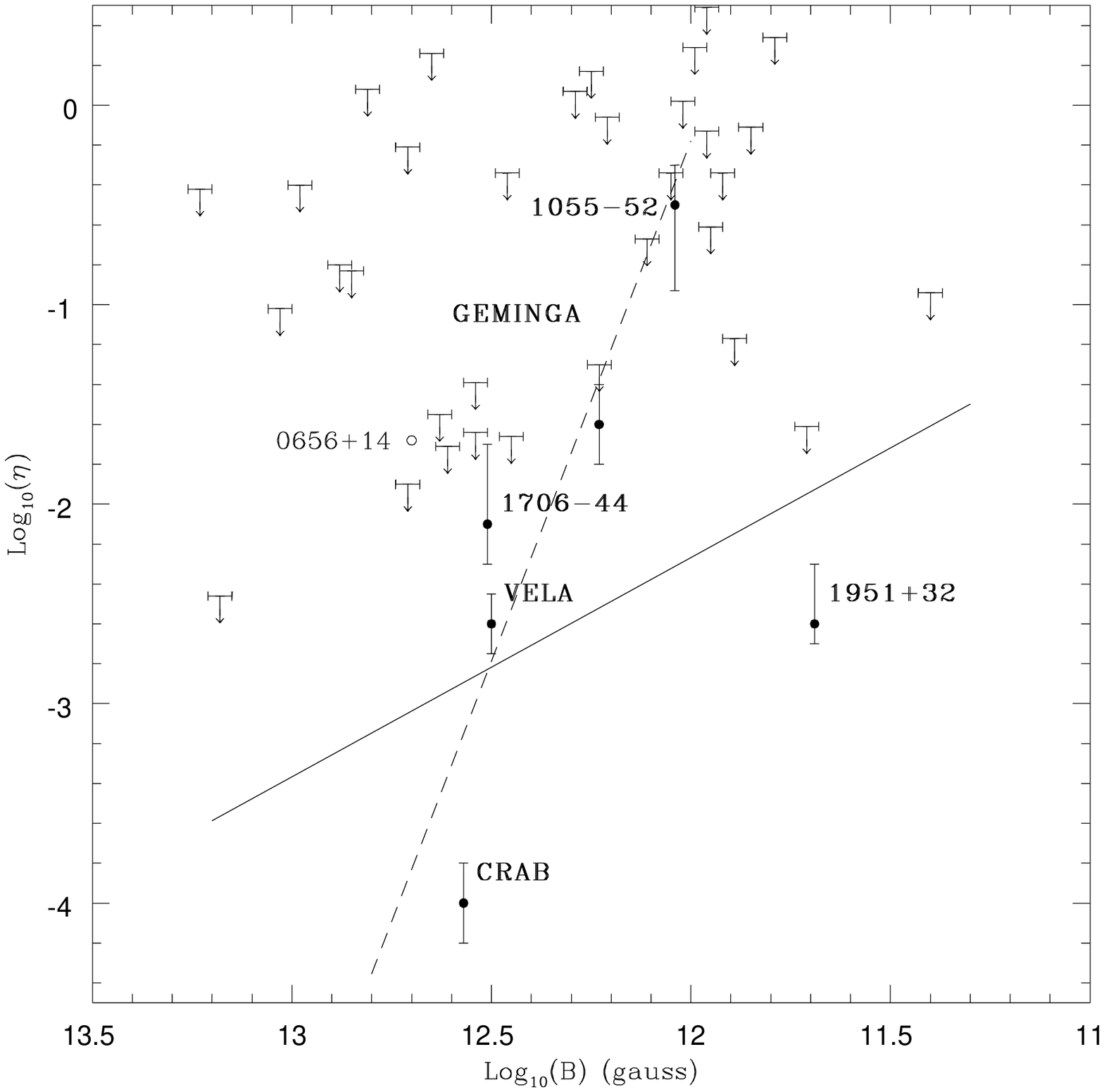,height=85mm}}
\centerline{\psfig{figure=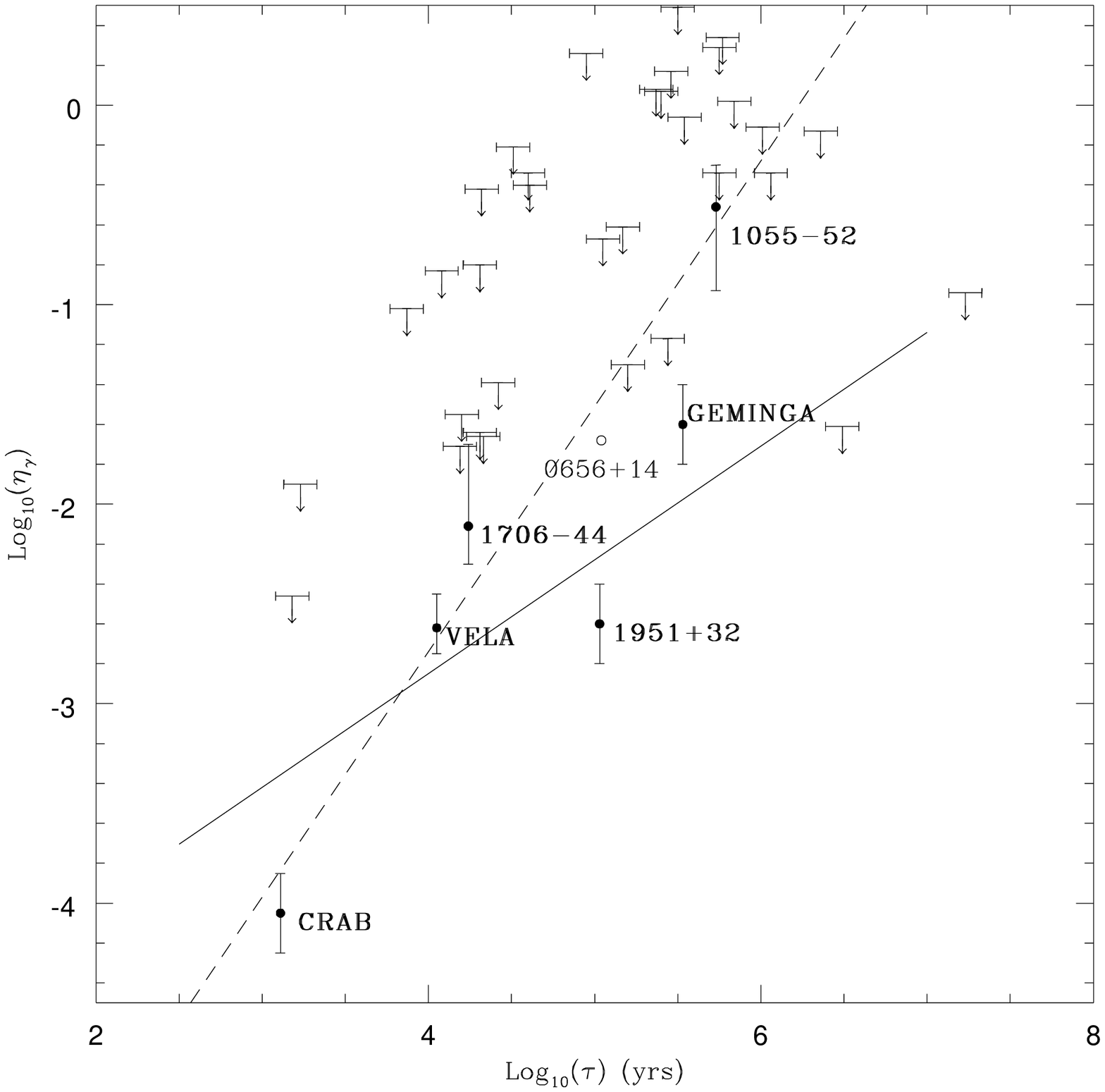,height=85mm}}
\caption{$\gamma$ efficiency as a function of B and $\tau $, for the
$\dot E / d^2$ selected sample; superimposed are linear fits of the first 5
detected points (dashed) and of the whole sample with the upper limits (solid).}

\label{Figure1}
\end{figure}

We thus conclude that a single linear fit of INS $\gamma $-ray efficiencies as a
function of any rotational parameter, with the possible exception of age, is not
acceptable in the EGRET energy band. As it was shown by Goldoni et al. (1995),
the bulk of the emission of middle-aged INSs falls in this energy range. The
lack of detections of lower-$B$ INSs in this band is a very important observational
fact, as it suggests that either the overall electromagnetic efficiency decreases,
or a great part of the energy loss fills up another region of the spectrum.

Is really the overall electromagnetic efficiency decreasing? In the frame of the
outer gap model (Cheng et al. 1986a,b) the answer could be yes.

In the polar cap models (Ruderman \& Sutherland 1975; Daugherty \& Harding 1982;
Arons 1983; Harding et al. 1993) the answer is not unique. A simple physical
explanation, concerning the number of emitted photons and the cutoff energy of
the $\gamma$-ray emission could be given.

Old pulsars (like PSR 0950+08 and PSR 1929+10) Have hard spectra at high energies,
with spectral indexes $\delta $ approaching unity and possibly becoming even less
than one. At the same time the high-energy cutoff of the emission grows, owing to
the reduced optical depth of the magnetosphere (this is true if the spectral break
is due to pair production rather than to curvature radiation cutoff). In this way
we could find $\gamma $-ray efficiencies very high without being observed by EGRET.

For instance, for $\delta =-0.9$ and $E_{break}=45~GeV$, PSR 1929+10 would have
$\eta_{\gamma}>0.3$ while being undetectable by EGRET. In fact its spectrum, combined
with the efficiency, would give at $E>100~MeV$ a flux $F\sim 6.6\times 10^{-9}~ph/cm^2/s$,
much lower than the Thompson et al. (1994) upper limit of $1.63\times 10^{-7}~ph/cm^2/s$.
If we instead look at the flux at $E>1~GeV$ the situation is better, thanks to the
lower value of the $\gamma $-ray background and the better instrument efficiency:
the expected flux is $F\sim~5\times 10^{-9}~ph/cm^2/s$, only about four times less than
the upper limit of $1.7\times 10^{-8}~ph/cm^2/s$. The same arguments hold for PSR 0950+08.

However this is not the case for PSR 1951+32, which is expected to be a harder $\gamma $-ray
emitter in this frame, with $\delta \sim -1.4$ instead of the observed
$\delta \sim -1.74\pm 0.11$. This is a question to be studied separately, owing to the peculiar
rotational parameters of this object which led White \& Stella (1988) to suggest it is a "recycled" pulsar.

Clearly the way to overcome present day difficulties is to develop new instruments
with higher effective areas in this energy range. For example,
there exists a SLAC project, called GLAST, which should have a $8000~cm^2$
effective area from 0.01 to $300~GeV$, and a limiting flux
$F_{min}=1.5\times10^{-10}~ph/cm^2/s$ between 1 and $300~GeV$ (Michelson 
1995).

\section {Conclusions}

We investigated the high-energy $\gamma $-ray emission of the 40 INSs
with large $\dot E/d^2$, all detected as radio pulsars with the exception
of Geminga. Due to the low number of detections with respect to non-detections,
the only way to correctly deal with these data is to include non-detections
as well as detections in our analysis. The presence of non-detections, in fact,
rules out the possibility to perform a linear fit between $\eta_{\gamma}$ and
any INS' rotational parameter.

A possible explanation of this phenomenon can be given in the framework
of outer gap and polar cap models. In the first case the $\gamma $-ray
emission is quenched for the majority of radio pulsars (Chen \& Ruderman 1993),
and the INS is no more a strong $\gamma $-ray emitter. In the second the
electromagnetic emission, still very strong, becomes harder, emitting the bulk
of its energy at $E>10~GeV$ thus becoming unobservable by current instruments.
Future instruments with good sensitivity at $E\sim 100~GeV$ currently being
studied, will address this question properly.
 
\acknowledgements One of us (P.G.) acknowledges useful comments by E.D. Feigelson on ASURV software

\end{document}